Self-gravitating clouds of generalized Chaplygin and modified anti-Chaplygin Gases

T.C. Lipscombe

JHUP, 2715 North Charles Street, Baltimore, MD 21218.

The Chaplygin gas has been proposed as a possible dark energy, dark matter candidate. As a working fluid in a Friedmann-Robertson-Walker universe, it exhibits early behavior reminiscent of dark matter, but at later times is more akin to a cosmological constant. In any such universe, however, one can expect local perturbations to form. Here we obtain the general equations for a self-gravitating relativistic Chaplygin gas. We solve these equations and obtain the mass-radius relationship for such structures, showing that only in the phantom regime is the mass-radius relationship large enough to be a serious candidate for highly compact massive objects at the galaxy core. In addition, we study the cosmology of a modified anti-Chaplygin gas. A self-gravitating cloud of this matter is an exact solution to Einstein's equations.

Sergei Chaplygin, in his studies of aerodynamics, introduced an equation of state that was, at the same time, straight forward yet exotic[1]. The pressure $P$ in a simple Chaplygin gas (SCG) is related to its density $\rho$ through:

$$P = -\frac{A}{\rho}. \qquad (1.1)$$

Its close cousin, the generalized Chaplygin gas (GCG) has the equation of state:

$$P = -\frac{A}{\rho^\alpha}. \qquad (1.2)$$

Clearly, the SCG corresponds to a GCG with α = 1. In the general case, one usually considers only those exponents for which $0 \leq \alpha \leq 1$. One physical reason for this restriction of the exponent is that polytropic self-gravitating clouds of gas can have infinite mass if α is outside this range. That said, Bertolami et al[2] have considered values of α greater than one, while Viala and Hovedt[3] looked at polytropes with negative exponents though, strictly speaking, these are anti-Chaplygin gases, which lack the minus sign in Eq.(2).

In this article, we consider gas clouds that consist of Chaplygin or ant-Chaplygin gases. First we extend the results of other workers who have studied simple and/or nonrelativistic Chaplygin gases. Then we look at the evolution of a universe that comprises of a modified anti-Chaplygin gas and show that a perturbation of such a cloud may form an exact solution of Einstein's equations.

## 1. The generalized Chaplygin Gas

The pressure in the GCG is negative, which suggests a connection to the cosmological constant. As noted in [2], if one does cosmology on a Friedmann-Robertson-Walker metric with the generalized Chaplygin gas, then conservation of energy requires that:

$$\rho = \left[ A + \frac{B}{a^{3(1+\alpha)}} \right]^{1/1+\alpha}. \tag{1.3}$$

Here, $a$ is the usual scale factor of the universe and $B$ is a positive constant. In the early universe, ρ behaves like nonrelativistic dark matter, whereas at later time density is approximately constant, therefore resembling a cosmological constant. As such, it may be the "missing link" between dark matter and dark energy[4].

Given the intriguing behavior of the generalized Chaplygin gas in cosmology, it is natural to explore the properties of a self-gravitating cloud of generalized Chaplygin gas in the relativistic regime. Bilić, Tupper, and Viollier[5] discussed an SGC in the so-called phantom-fluid regime, where $P+\rho <0$. They found a gravastar solution to Einstein's field equation. In a similar fashion, Bertolami and Paramos[6] (BP) studied a self-gravitating cloud comprised of a GCG, but focused only on its non-relativistic aspects.

Here we extend the results of BTV from the SGC to the GCG, and to extend the results of BP to the relativistic realm. By so doing, we will establish two different mass-radius relations for a self-gravitating cloud of Chaplygin gas. This is important: if we believe that the generalized Chaplygin gas is truly a dark-matter candidate, then we need to look at stable configurations of condensed clouds of the gas. The mass-radius relation is important in determining whether a GCG might truly be a candidate to explain dark matter.

The starting point for the description of a self-gravitating sphere of a perfect GCG is the Tolman-Oppenheimer-Volkoff (OV) equation -- the relativistic counterpart of the equation of hydrostatic support. For a perfect fluid, the OV equation can be written as[7]:

$$\frac{dP}{dr} = -\frac{G(P+\rho)[m(r)+4\pi r^3 P]}{r[(r-2Gm(r)]} \tag{1.4}$$

wherein $G$ is Newton's gravitational constant and $m(r)$ is given by:

$$\frac{dm}{dr} = 4\pi\rho r^2. \tag{1.5}$$

Outside the cloud of gas, the metric must go over to the usual Schwartzschild metric, a spherically symmetric static solution to Einstein's vacuum equations. Equations (4) and (5) are supplemented by the GCG equation of state, to form three equations with three unknowns. By use of the equation of state, can replace $dP/dr$ in Eq.(4) by $d\rho/dr$. Doing so, we obtain:

$$\frac{d\rho}{dr} = -\frac{G\rho(\rho^{\alpha+1}-A)[m(r)-\frac{4\pi A r^3}{\rho^\alpha}]}{A\alpha r[(r-2Gm(r)]} \tag{1.6}$$

Given A, α, and the appropriate boundary conditions, Eqs (5) and (6) form a closed system that can be integrated numerically.

We simplify Eq.(6) by insisting that it deals with the phantom regime. The term $(\rho^{\alpha+1} - A)$ will be negative on the surface of the cloud, given that the density there will be zero. The maximum value of the density occurs at center of the cloud. Thus if $\rho_c^{\alpha+1} < A$, this term will always be negative. This same condition was found, for α = 1, by BTV. We know, from Eq.(5), that $dm/dr < 4\pi\rho_c r^2$. Thus the second term in the numerator of Eq.(6) is negative provided that $\rho_c^{\alpha+1} < 3A$. If this condition holds, the entire RHS of Eq.(6) will be negative and so the density in the gas cloud will indeed decrease outwards, as we would expect for a realistic gravitational object.

Following BTV, we assume that $\rho_c^{\alpha+1} \ll A$ and go to the nonrelativistic limit, Eq.(6) goes over to a far simpler form, namely:

$$\rho^{\alpha-1}\frac{d\rho}{dr} = -4\pi G\alpha r \qquad (1.7)$$

whose solution is:

$$\rho^\alpha = 2\pi A G \alpha^2 R_0^2 (1 - \frac{r^2}{R_0^2}). \qquad (1.8)$$

The central density of the cloud is therefore $\rho_c = (2\pi A G \alpha^2 R_0^2)^{1/\alpha}$.

We can determine the mass of the cloud and, more generally, the mass-radius relationship for a self-gravitating GCG. The mass is:

$$M = 4\pi \rho_c \int_0^{R_0} \left(1 - \frac{r^2}{R_0^2}\right)^{1/\alpha} r^2 dr.  \tag{1.9}$$

As a consequence, the mass-radius relationship is:

$$M = 2\pi R_0^3 \left(2\pi A G \alpha^2 R_0^2\right)^{1/\alpha} B(\frac{3}{2}, 1 + \frac{1}{\alpha})  \tag{1.10}$$

in which $B(a,b)$ is the beta function. For a self-gravitating GCG, Eq.(10) predicts that $M \propto R_0^{3+2/\alpha}$. For the SCG, the mass varies as the fifth power of the radius, a result obtained by BTV. Note that this opens up the bright prospect that phantom gravastars could be the dark supermassive objects at the galactic core. For small values of α, even a relatively compact GCG gravastar will be massive.

We need to show that such an object can be stable. Collapsing shells of Chaplygin gas were discussed by Mann and Oh[8]. A GCG in the phantom regime can only be stable if contained by a thin shell whose surface density is σ and whose surface tension is θ and held in by an associated potential, $V(R)$. Visser and Wiltshire[9], show that such gravastars are stable if:

$$4\pi\sigma = -\frac{1}{G}\left[\left[\sqrt{1 - \frac{2Gm(R)}{R}} - 2V\right]\right]  \tag{1.11}$$

and

$$4\pi(\sigma-\theta) = \left[\left[\frac{-R^2 V'(R) + Gm(R) - 4\pi GR^3 \rho}{GR^2\sqrt{1-\frac{2Gm(R)}{R}-2V(R)}} + \frac{4\pi R(p+\rho)\sqrt{1-\frac{2Gm(R)}{R}}-V(R)}{1-\frac{2Gm(R)}{R}}\right]\right] \quad (1.12)$$

Here [[Q]] denotes the difference in the quantity Q between its value just above the cloud's surface and a point just below. These formulas ultimately derive from Israel's junction condition[10] for the extrinsic curvature, and apply equally well to wormholes[11,12].

Consider the static shell, for which the potential $V \equiv 0$. Equation (1.35) reduces to:

$$4\pi G\sigma_{static} = \sqrt{1-\frac{2Gm(R)}{R}} - \sqrt{1-\frac{2GM}{R_0}} \quad (1.13)$$

Likewise the equation for the surface tension becomes:

$$4\pi(\sigma_{static} - \theta_{static}) = \frac{M}{\sqrt{1-\frac{2GM}{R}}} - \frac{m(R) - (4\pi R^3 A/\rho^\alpha)}{\sqrt{1-\frac{2Gm(R)}{R}}}. \quad (1.14)$$

Both of these equations extend the results for the simple Chaplygin gas to the general case. With these two equations, one can solve for σ and θ, thus providing an equation of state for the material of which the shell must consist.

To explore further, assume that the density-radius function is given by Eq(1.14). In addition, we assume that condition (1.11) holds. Also, we take it that $2GM/R_0 \ll 1$.

Evaluate the various quantities on $R_0$ and $R_0(1-y)$. The mass contained within $R_0$ is the mass of the self-gravitating gas, $M$. The mass $m(R)$ contained within a radius $R$ is:

$$m(R) = 2\pi R_0^3 \left(2\pi A G \alpha^2 R_0^2\right)^{1/\alpha} B\left(\frac{R^2}{R_0^2}; \frac{3}{2}, 1+\frac{1}{\alpha}\right) \qquad (1.15)$$

where $B(x; a,b)$ is the incomplete beta function[13].

Expand the incomplete beta function as a function of $y$. The easiest way to do is by use of the regularized incomplete beta function, $I_z(a,b)$, which has two useful properties. First, $I_z(a,b) = 1 - I_{1-z}(b,a)$. Second,

$$I_z(a,b) = z^a \left[\frac{1}{a} + ...\right]. \qquad (1.16)$$

Substitution in Eq.(1.30) gives, to leading order in y:

$$\sigma = \frac{\rho_c R_0 \alpha (2y)^{1+1/\alpha}}{2(1+\alpha)} = \frac{\alpha R_0 y}{1+\alpha}\left(4\pi A G \alpha^2 R_0^2 y\right)^{1/\alpha}. \qquad (1.17)$$

When α = 1, this expression reduces to:

$$\sigma = \rho_c R_0 y^2 = 2\pi A G R_0^3 y^2, \qquad (1.18)$$

which is exactly the result obtained for a simple Chaplygin gas by BTV.

By the same method, to leading order in $y$, the tension is given by

$$\theta = -\frac{1}{8\pi G \alpha^2 R_0 y} \qquad (1.19)$$

Again, this reduces to the result of BTV when α = 1.

Together, Equations (1.37) and (1.39) constitute an equation of state that links $\theta$ to $\sigma$. The pressure $P = -\theta$ so that

$$P = \frac{1}{4\pi G \alpha^2 R_0} \left[ \frac{\rho_c R_0 \alpha}{2\sigma(1+\alpha)} \right]^{\alpha/(1+\alpha)} = const. \frac{1}{\sigma^{\alpha/(1+\alpha)}} \qquad (1.20)$$

which is the equation of state for an *anti*-Chaplygin gas. When $\alpha=1$, the equation reduces to:

$$P = \left( \frac{AR_0}{32\pi G \sigma} \right)^{1/2} \qquad (1.21)$$

which again is the result of BTV.

We now look at the mass-radius relationship in different types of self-gravitating GCGs in a different regime. Clearly the GCG is closely related to a polytropic equation of state. Bertolami and Páramos showed that if we define:

$$r = \left[ \frac{A}{4\pi G \rho_c^{(2+\alpha)}} \frac{\alpha}{1+\alpha} \right]^{1/2} \xi, \qquad (1.22)$$

and $\rho = \rho_c \theta^n$, then:

$$\frac{1}{\xi^2} \frac{d}{d\xi} \left( \xi^2 \frac{d\theta}{d\xi} \right) = +\theta^n \qquad (1.23)$$

in which the power $n = -1/(1+\alpha)$.

This GCG equation might be called the imaginary Lane-Emden equation, for if $\xi \to i\xi$ in Eq.(1.23), we obtain Eq.(1.18). But the mass-radius relationship for a polytrope of index $n$ is $M \propto R^{(3-n)/(1-n)}$. This still holds for the GCG, so that $M \propto R^{(4+3\alpha)/(2+\alpha)}$.

For the simple Chaplygin gas, $M \propto R^{7/3}$. When $\alpha$ is small, $M \propto R^2$. Note that this expression for the mass has a far smaller exponent than that in the phantom GCG. That is to say, if a compact Chaplygin gas is to be plausible as a dark-matter candidate, then it needs to be in the phantom regime.

There are singular solutions for the GCG, which we can regard as the limiting case of high central density. (Nonrelativistic) hydrostatic equilibrium requires:

$$\frac{1}{r^2}\frac{d}{dr}\left(\frac{r^2}{\rho}\frac{dP}{dr}\right) = -4\pi G\rho \tag{1.24}$$

Substituting the expression $P = -A/\rho^\alpha$, setting $\rho = \rho_c/r^b$, and balancing the powers of $r$ that result leads to:

$$\rho = \frac{\rho_c}{r^{2/(2+\alpha)}} \tag{1.25}$$

where the central density is:

$$\rho_c^{2+\alpha} = \frac{A}{2\pi G}\frac{4\alpha + 3\alpha^2}{(2+\alpha)^2}. \tag{1.26}$$

The mass-radius relation for this type of cloud is also $M \propto R^{(4+3\alpha)/(2+\alpha)}$, the same dependence as solutions generated from the imaginary Lane-Emden equation. Note that when $\alpha = 1$, this reduces to:

$$\rho(r) = \left(\frac{7A}{18\pi G r^2}\right)^{1/3} \tag{1.27}$$

which is exactly the equation derived by BTV.

## 2. Self-gravitating Anti-Chaplygin gas clouds

For a gravastar of Chaplygin gas to exist, there has to be a thin shell of exotic matter that, as we have seen above, consists of an anti-Chaplygin gas. Thus far, such a gas has not been well studied in the literature. The equation of state for an anti-Chaplygin gas is:

$$P = \frac{B}{\rho^\beta} \tag{1.28}$$

In other words, an anti-Chaplygin gas is a negative-index polytrope. As such, in the nonrelativistic limit, it obeys the Lane-Emden equation.

Here, though, we introduce a *modified* anti-Chaplygin gas. To do so, we replace Eq.(28) with a more general form:

$$P = \frac{B}{\rho^\beta} - Y\rho. \tag{1.29}$$

This can be thought of as an anti-Chaplygin gas that also has an "inflation" term. The pressure term is similar to that proposed by Bedran, Soares, and Araujo[14]. If we substitute this expression into the Tolman-Oppenheimer-Volkoff equation (4) we can find an *exact* solution. If we require the density to go as a power of the radius, then an exact solution exists[15] when:

$$P = \frac{3\beta + 1}{4\rho^\beta} - \beta\rho; \quad \rho = \rho_c r^{2/\beta} \tag{1.30}$$

for which the central density is

$$\rho_c = \left[\pi\left(\frac{3\beta+1}{2}\right)\right]^{1/\beta} \qquad (1.31)$$

This structure has a number of intriguing properties. In a Chaplygin gas cloud, the density falls off with radius while the pressure increases. Thus, at the surface, the pressure is infinite. For the modified anti-Chaplygin gas, the opposite holds true. The pressure falls of with radius, whereas the density increases.

From Eq(30) and (31), we know how the density varies as a function of radius. Thus we can find the mass of the cloud, which is:

$$M = \frac{(4\pi\beta)}{(3\beta+2)}\left[\frac{(3\beta+1)\pi}{2}\right]^{1/\beta} R^{3+2/\beta} \qquad (1.32)$$

If, however, we require that the surface of the cloud be the radius where the pressure is zero, we obtain:

$$\frac{3\beta+1}{4\rho_c^{\beta+1}} = R^{2+2/\beta}. \qquad (1.33)$$

Combining these equations gives the mass-radius relationship as:

$$M = \left(\frac{2\beta}{3\beta+2}\right)R, \qquad (1.34)$$

which is a linear relation between mass and radius.

## 3. FRW cosmology with an anti-Chaplygin gas

The energy density in FRW cosmology obeys[16]:

$$\dot{\rho} = -3(p+\rho)\frac{\dot{a}}{a} \qquad (1.35)$$

where $a$ is the scale factor. For this particular fluid, we find:

$$\rho = \frac{1}{(1-\beta)}\left[\frac{K}{a^{3(\beta^2-1)}} - \frac{3\beta+1}{4}\right]^{1/(1+\beta)} \qquad (1.36)$$

where $K$ is a constant of integration. Note that this gas leads to a finite universe, for the energy density falls to zero when the scale factor is:

$$a_{max} = \left(\frac{4K}{3\beta+1}\right)^{1/3(\beta+1)}. \qquad (1.37)$$

If the scale factor is 1 when the energy density is $\rho_0$, then we can express the maximum scale factor as:

$$a_{max} = \left[1 + \frac{1-\beta}{B}\rho_0^{1+\beta}\right]^{1/3(1-\beta^2)} \qquad (1.38)$$

Early in the universe, when the scale factor is small, $\rho \sim 1/a^{3(1-\beta)}$. Thus if β=0, we have a universe comprised of dust. We can go further by looking at the dynamics of such a universe. We know that[17]:

$$\dot{a}^2 = -k + \frac{8\pi\rho}{3}a \qquad (1.39)$$

where $k$ is 1, 0, or -1. From this, we know that if the universe is flat, so that $k=0$, then the universe is "stationary" when $\rho a=0$. This occurs when $\rho=0$, which is when $a = a_{max}$. Thus, in a flat Friedmann model filled with a modified anti-Chaplygin gas, the universe reaches a finite radius in a finite time.

For $k=1$, the situation is more complex. The universe ceases to expand when

$$\rho a = \frac{3\pi}{8} \tag{1.40}$$

This is solved when:

$$\frac{(1-\beta)\rho_0^{1+\beta} + B}{(1-\beta)a^{3(1-\beta^2)}} - \left(\frac{3}{8\pi a}\right)^{1+\beta} - \frac{B}{1-\beta} = 0 \tag{1.41}$$

or equivalently, when

$$\frac{B}{(1-\beta)}\left[\left(\frac{a_{max}}{a}\right)^{3(1-\beta^2)} - 1\right] = \left(\frac{3}{8\pi a}\right)^{1+\beta} \tag{1.42}$$

The general solution is difficult to write down. A simple case, though, is when the powers of $a$ match, i.e., when $3(1-\beta^3) = 1+\beta$, which happens when $\beta = 2/3$. In that case, the solution is:

$$a^{5/3} = a_{max}^{5/3} - \frac{4}{9}\left(\frac{3}{8\pi}\right)^{5/3} \tag{1.43}$$

This means that the expansion will halt before the energy density falls to zero. Other straightforward solutions can be obtained when $3(1-\beta^2) = 2(1+\beta)$ or $6(1-\beta^2) = (1+\beta)$, i.e., when $\beta=1/3$ and $1/6$ respectively. These results follow those of Setare[18]

## 4. Conclusions

In this article, we examined the behavior of self-gravitating gases that comprise of a Chaplygin fluid. We have found that, in the phantom regime, solutions to the relativistic equations exist that are stable. The mass of these objects goes as a high power of their radius. As such, they are good candidates for supermassive gravitating objects in the galactic center. We have also shown that nonrelativistic generalized Chaplygin gases have masses that depend on a far lower power of their radius, and are less attractive as dark-matter candidates. In all cases, results from the simple Chaplygin gas have carried over to the general Chaplygin gas.

In addition, we have introduced a modified anti-Chaplygin gas. A universe filled with this gas will evolve like matter in the early stages, but eventually the expansion will halt, shortly before the energy density would have dropped to zero. If, during the expansion, there are small perturbations that collapse to form isolated gravitational structures, we show that these are exact solutions of Einstein's equations. From this solution we have discovered a linear mass-radius relationship.

---

[1] S. Chaplygin, Sci. Men. Moscow Univ. Math. **21** 1 (1904).

[2] O. Bertolami, A.A. Sen, S. Sen, and P.T. Silver, Mon. Not. R. Astron. Soc. **353**, 329 (2004); A.A. Sen and Robert J. Scherrer, Phys. Rev. D **72**, 063511 (2005)

[3] Y.P. Viala and G. Horvedt, Astron. & Astrophys. **33** 195 (1974).

[4] M. Heydari-Fard and H.R. Sepangi, Phys. Rev. D. **76** 104009 (2007).